\documentclass[aps,amsfonts,amsmath,amssymb,nofootinbib,twocolumn,superscriptaddress]{revtex4}

\usepackage{braket}
\usepackage{graphics}
\usepackage{hyperref}
\usepackage{epsfig}
\usepackage{color}
\usepackage{verbatim}
\usepackage{soul}

\usepackage[normalem]{ulem}
\usepackage{multirow}
\usepackage{bm}

\usepackage{pifont}
\usepackage{amssymb}
\usepackage{amsmath}

\begin{document}

\bibliographystyle{naturemag}

\title{Revealing Rotational Symmetry Breaking Charge-density Wave Order in Kagome Superconductor (Rb, K)V$_3$Sb$_5$ by Ultrafast Pump-probe Experiments}

\author{Qinwen Deng}
\affiliation{Department of Physics and Astronomy, University of Pennsylvania, Philadelphia, Pennsylvania 19104, U.S.A}
\author{Hengxin Tan}
\affiliation{Department of Condensed Matter Physics, Weizmann Institute of Science, Rehovot, Israel}
\author{Brenden R. Ortiz}
\affiliation{Materials Department, University of California Santa Barbara, Santa Barbara, California
93106, U.S.A.}
\affiliation{Materials Science and Technology Division, Oak Ridge National Laboratory, Oak Ridge, Tennessee 37831, United States}
\author{Stephen D. Wilson}
\affiliation{Materials Department and California Nanosystems Institute, University of California Santa Barbara, Santa Barbara, California
93106, U.S.A.}
\author{Binghai Yan}
\affiliation{Department of Condensed Matter Physics, Weizmann Institute of Science, Rehovot, Israel}
\author{Liang Wu}
\email{liangwu@sas.upenn.edu}
\affiliation{Department of Physics and Astronomy, University of Pennsylvania, Philadelphia, Pennsylvania 19104, U.S.A}

\date{\today}

\begin{abstract}
The recently discovered Kagome superconductor AV$_3$Sb$_5$ (where A refers to K, Rb, Cs) has stimulated widespread research interest due to its interplay of non-trivial topology and unconventional correlated physics including charge-density waves (CDW) and superconductivity. The essential prerequisite to understanding the microscopic mechanisms of this complex electronic landscape is to unveil the configuration and symmetry of the charge-density wave order. As to now, little consensus has been made on what symmetry is broken. Herein, we clarify the microscopic structure and symmetry breaking of the CDW phase in RbV$_3$Sb$_5$ and KV$_3$Sb$_5$ by ultrafast time-resolved reflectivity. Our approach is based on extracting coherent phonon spectra induced by three-dimensional CDW and comparing them to calculated phonon frequencies via density-functional theory. The combination of these experimental results and calculations provides compelling evidence that the CDW structure of both compounds prevailing up to T$_{\text{CDW}}$ is the 2 $\times$ 2 $\times$ 2 staggered inverse Star-of-David pattern with interlayer $\pi$ phase shift, in which the six-fold rotational symmetry is broken. These observations thus corroborate six-fold rotational symmetry breaking throughout the CDW phase of RbV$_3$Sb$_5$ and KV$_3$Sb$_5$. 
\end{abstract}

\maketitle 

When the energy scale of interaction between electrons is on the same order of the electron kinetic energy, a wealth of unconventional strongly correlated phases  emerge. In this regard, the newly discovered Kagome materials AV$_3$Sb$_5$ (A = K, Rb, Cs) have been at the forefront of this field, providing a new platform to study the interaction between a plethora of exotic correlated phases. This group of materials host a unique phase transition into charge-density wave (CDW) order with T$_{\text{CDW}}$ = 78-102 K followed by superconductivity at low temperatures with T$_{\text{c}}$ = 0.9-2.5 K\cite{ortiz2019new, ortiz2020cs, ortiz2021superconductivity, yin2021superconductivity}. The CDW order sets the stage for a cascade of intertwined symmetry breaking orders developing concomitantly or subsequently with the CDW transition, including a possible orbital flux phase\cite{mielke2022time, jiang2021unconventional, xing2024optical, guo2022switchable, xu2022three, park2021electronic, feng2021chiral, denner2021analysis, guo2024correlated, christensen2022loop, tazai2024drastic, saykin2023high, farhang2023unconventional}, electronic nematicity\cite{nie2022charge, li2023unidirectional, xiang2021twofold, chen2021roton, wulferding2022emergent, asaba2024evidence, liu2024absence}
, and superconductivity. Therefore, determining the exact microscopic configuration and symmetry of the CDW order is of top significance to establish the foundation of understanding these unconventional many-body effects. 

\begin{figure}
    \centering
    \includegraphics[width=\columnwidth]{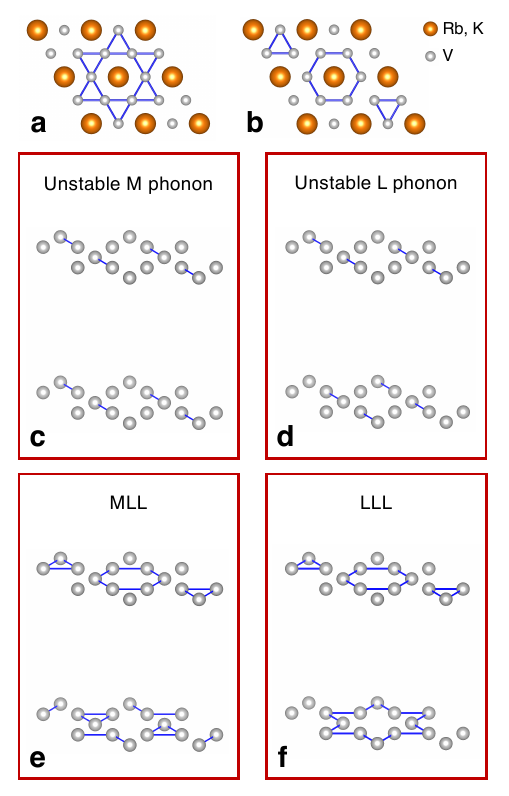}
    \caption{\textbf{Unstable phonon modes and CDW distortions in RbV$_3$Sb$_5$ and KV$_3$Sb$_5$.} (a) 2 $\times$ 2 $\times$ 1 SD. (b) 2 $\times$ 2 $\times$ 1 ISD. (c) Distortion corresponding to the unstable M phonon. (d) Distortion corresponding to the unstable L phonon. (e) ISD + ISD with interlayer $\pi$-phase shift via combination of one M and two L unstable phonons (MLL). (f) SD + ISD without interlayer phase shift via combination of three L unstable phonons (LLL). From (a) to (f) The V atoms are shown in gray. In (a) and (b) the alkali metal atoms (Rb or K) are shown in orange, and from (c) to (f) only the vanadium atoms are shown for simplicity. The lines connecting V atoms indicate shorter bonds. Note that the structure in (a), (b) and (f) keeps the $D_{6h}$ symmetry while the structure in (e) breaks the six-fold rotational symmetry down to two-fold (point group $D_{2h}$). }
    \label{fig1}
\end{figure}

AV$_3$Sb$_5$ host a Kagome lattice of vanadium atoms with space group P6/mmm in the normal phase above T$_{\text{CDW}}$. The pristine phases exhibit structural instability, with unstable phonon modes at the M and L points of the Brillouin zone in all 3 compounds\cite{tan2021charge}. Note there are three symmetry equivalent M points and
three symmetry equivalent L points, thus there are three M modes and three L modes in total. We show the distortion patterns of the unstable M phonon and the unstable L phonon respectively in Fig. \ref{fig1}c, d\cite{christensen2021theory, kang2022charge, ning2024dynamical}. Other symmetry equivalent M and L modes have similar distortion patterns. The soft phonons at the M point can be characterized by breathing modes of V atoms forming either a Star-of-David (SD) or inverse Star-of-David (ISD) structure\cite{park2021electronic, tan2021charge, christensen2021theory}, both of which exhibit an in-plane 2 $\times$ 2 supercell with preserved six-fold rotational symmetry. Then, combining three Q wavevectors of unstable M and L phonons, one can construct various 2 $\times$ 2 $\times$ 2 3D CDW orders\cite{tan2021charge, christensen2021theory, kang2022charge}. For example, by condensing one M and two L unstable phonons on three \textbf{Q} directions (MLL), one realizes the ISD + ISD with interlayer $\pi$ phase shift order (Fig. \ref{fig1}e). Thus the ISD + ISD with interlayer $\pi$ phase shift order is denoted as MLL. Via condensing three L unstable phonons on three \textbf{Q} directions (LLL), one realizes the alternative SD + ISD without interlayer $\pi$ phase shift order (Fig. \ref{fig1}f). Thus the alternative SD + ISD without interlayer $\pi$ phase shift order is denoted as LLL. Despite elaborate research, consensus on the precise real-space structure of the CDW state among AV$_3$Sb$_5$ has not been reached. Due to the complex energetic landscape, there are various competing CDW real-space structures with similar energy scale\cite{tan2021charge, christensen2021theory, ning2024dynamical}. In RbV$_3$Sb$_5$ and KV$_3$Sb$_5$, which are our focus in this study, several proposed CDW phases are: (i) 2 $\times$ 2 $\times$ 1 ISD \cite{zhang2024nmr, kato2022three} in which $C_6$ is preserved (Fig. \ref{fig1}b), (ii) ISD + ISD with interlayer $\pi$ phase shift\cite{kautzsch2023structural, kang2022charge, frassineti2023microscopic}(MLL. Also called staggered inverse Star-of-David state) in which $C_6$ is broken down to $C_2$ (Fig. \ref{fig1}e), and (iii) alternative SD + ISD without interlayer $\pi$ phase shift\cite{subires2023order, hu2022coexistence, kato2023surface} (LLL) in which $C_6$ is preserved (Fig. \ref{fig1}f). Here $C_6$ and $C_2$ refer to six-fold and two-fold rotational symmetry.  Evidently, there is still intense dispute on the exact rotational symmetry of the CDW structure in AV$_3$Sb$_5$.

In this study, we investigate ultrafast phonon dynamics in RbV$_3$Sb$_5$ and KV$_3$Sb$_5$ using optical pump-probe technique to evaluate the proposed CDW structures in these two compounds. By extracting coherent phonon spectra from the measurement of time-resolved reflectivity and comparing to density-functional theory (DFT) calculations, we determine the CDW structure to be most aligned with the 2 $\times$ 2 $\times$ 2 staggered inverse Star-of-David pattern with interlayer $\pi$ phase shift. Our work thus confirms $C_6$ rotational symmetry breaking onset at T$_{\text{CDW}}$, bringing forth new understandings of CDW order in the AV$_3$Sb$_5$ family.

\begin{figure*}
    \centering
    \includegraphics[width=17.2cm]{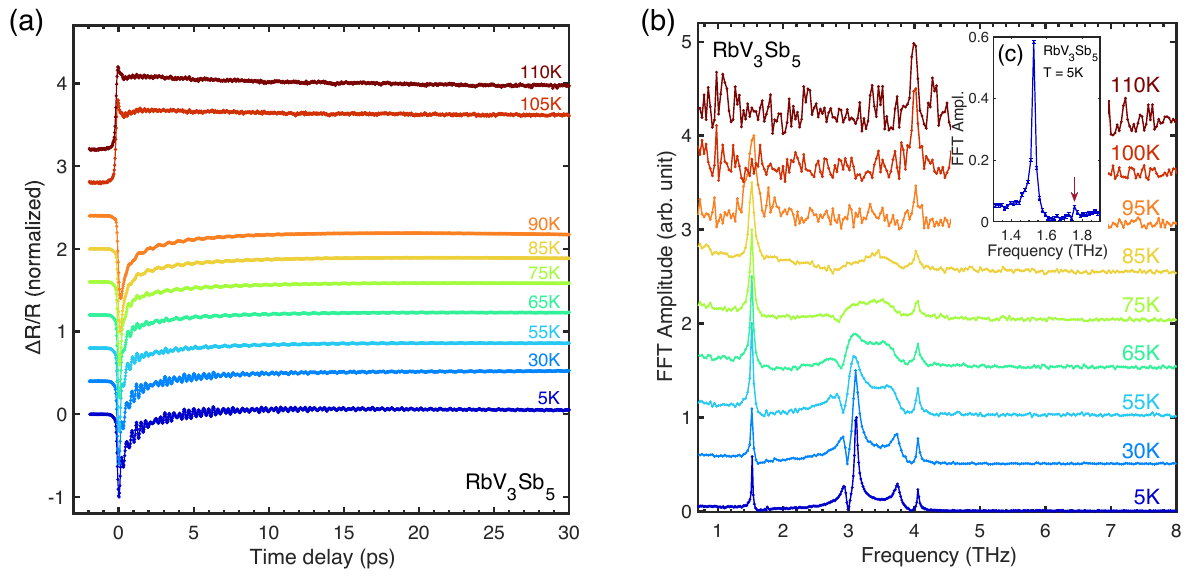}
    \caption{\textbf{Evolution of the coherent phonon spectrum in RbV$_3$Sb$_5$ vs. temperature. } (a) Time-resolved reflectivity curves $\Delta$R/R in the temperature range of 5 K – 110 K across the CDW transition temperature. (b) Amplitudes of Fourier transforms of coherent phonon oscillations in $\Delta$R/R time traces after subtracting the decay background. Inset (c) shows the weak 1.75 THz mode. Curves in (a) and (b) are offset for clarity.  
    }
    \label{fig2}
\end{figure*}

\begin{figure*}
    \centering
    \includegraphics[width=17.2cm]{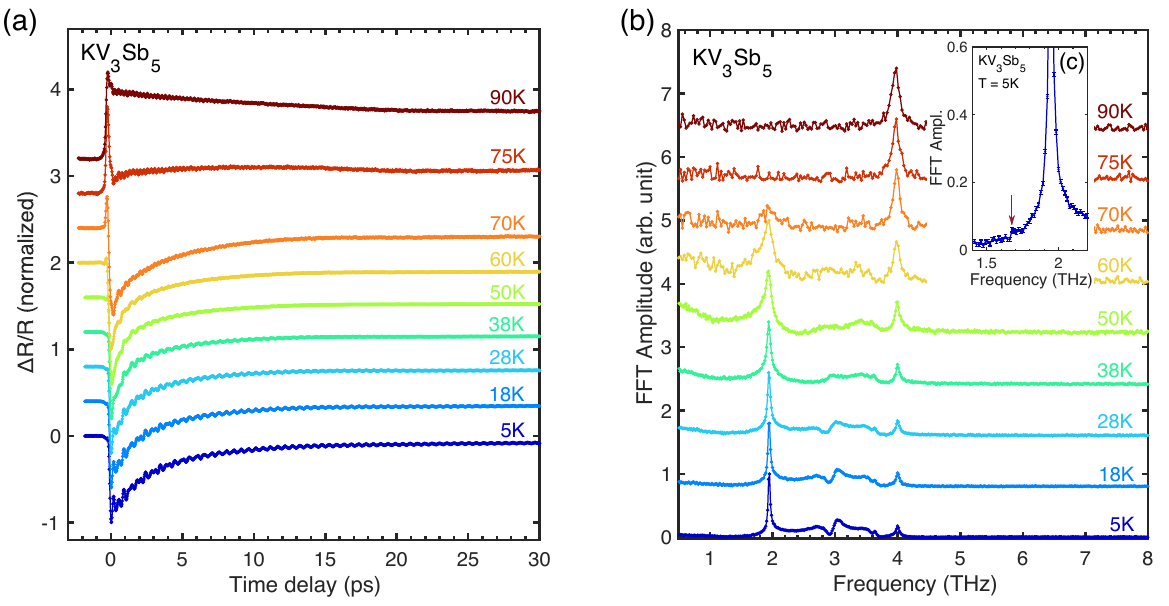}
    \caption{\textbf{Temperature-dependent coherent phonon spectroscopy in KV$_3$Sb$_5$. } (a) Time-resolved reflectivity curves $\Delta$R/R at different temperatures across the CDW transition temperature. Each curve is normalized to its peak value. (b) Amplitudes of Fourier transforms of coherent phonon oscillations in (a) after subtracting the decaying background. Inset (c) shows the weak 1.68 THz mode. Curves in (a) and (b) are offset for clarity. }
    \label{fig3}
\end{figure*}

\begin{figure*}
    \centering
    \includegraphics[width=17.2cm]{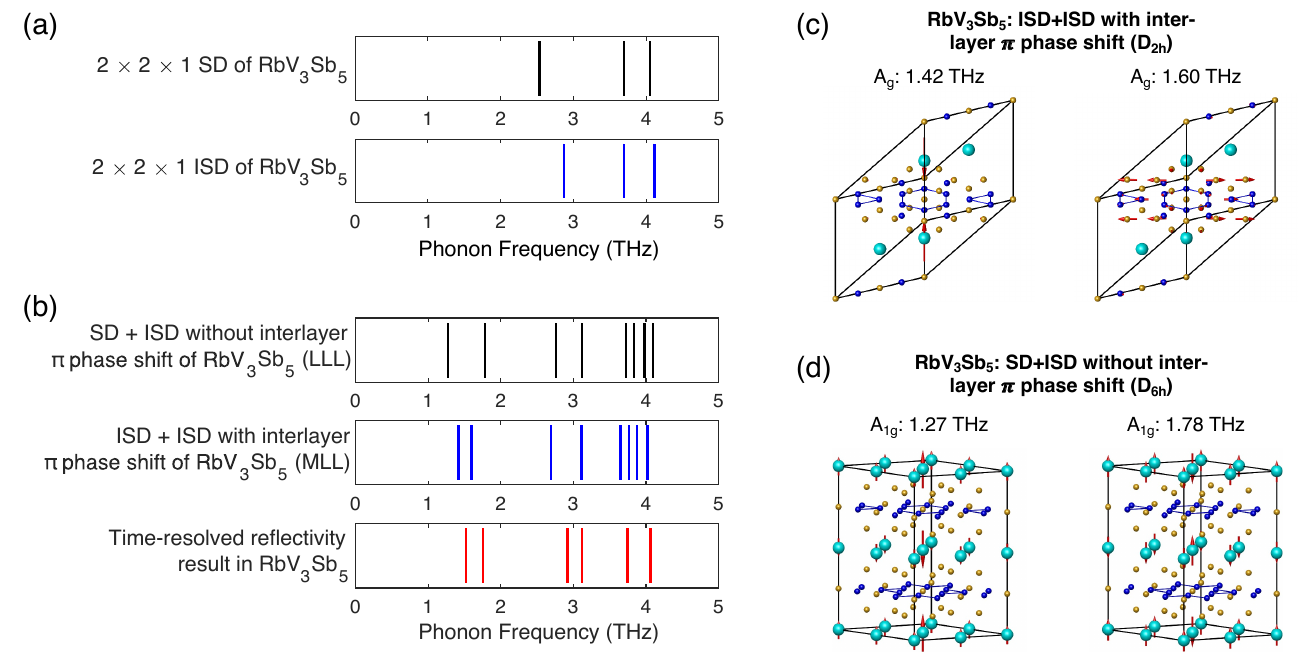}
    \caption{\textbf{Comparison of the phonon spectrum of RbV$_3$Sb$_5$ calculated by DFT and measured by time-resolved reflectivity at T = 5 K. } (a) DFT-calculated $A_{1g}$ phonon frequencies in the SD and ISD CDW phases in RbV$_3$Sb$_5$. (b) The calculated $A_{1g}$ ($A_g$) Raman mode frequencies in the SD+ISD without interlayer $\pi$ phase shift (ISD+ISD with interlayer $\pi$ phase shift) state and comparison with the time-resolved reflectivity results. The vertical lines in both figures denote the frequency of the measured phonon peak or the fully-symmetric $A_{1g}$ ($A_g$) modes in DFT calculation. (c) The DFT-calculated oscillation pattern of the fully-symmetric Raman active modes in ISD + ISD with interlayer $\pi$ phase shift (MLL) in RbV$_3$Sb$_5$ that are near our detected 1.53 and 1.75 THz modes via time-resolved reflectivity, respectively. The 1.42 and 1.60 THz $A_g$ modes are the lowest and second-lowest frequency mode in the $A_g$ phonon spectrum of MLL phase in (b). (d) The DFT-calculated oscillation pattern of the fully-symmetric Raman active modes in SD + ISD without interlayer $\pi$ phase shift (LLL) CDW state in RbV$_3$Sb$_5$ that are near our detected 1.53 and 1.75 THz modes via time-resolved reflectivity, respectively. The 1.27 and 1.78 THz $A_{1g}$ modes are the lowest and second-lowest frequency mode in the $A_{1g}$ phonon spectrum of LLL phase in (b). In (c) and (d), the Rb atoms are shown in cyan, the V atoms are shown in blue, and the Sb atoms are shown in yellow. 
    }
    \label{fig4}
\end{figure*}

Single crystals of RbV$_3$Sb$_5$ and KV$_3$Sb$_5$ were synthesized using the self-flux method\cite{ortiz2019new, ortiz2021superconductivity}. We perform time-resolved reflectivity measurements on freshly cleaved (001) surfaces of RbV$_3$Sb$_5$ and KV$_3$Sb$_5$ single crystals. We use an ultrafast fiber laser system for all the measurements, with a repetition rate of 80 MHz and pulse duration of 100 fs. The pump wavelength is 1560 nm and probe wavelength is 780 nm, with both having a fluence less than 10 $\mu$J/cm$^2$. The pump pulse intensity is modulated at a frequency of 84 kHz using a photo-elastic modulator (PEM) and a pair of linear polarizers. We use a balanced detection method to improve the signal to noise ratio. Both pump and probe beams are at normal incidence and are focused by an objective lens to achieve a spot size of $\approx$10 $\mu$m in diameter\cite{gray2024time}. 

Fig. \ref{fig2}a shows the time-resolved reflectivity $\Delta$R/R data measured on RbV$_3$Sb$_5$ and its evolution with temperature. Above the CDW transition temperature T$_{\text{CDW}}$, a single phonon oscillation can be observed. In contrast, multiple phonon oscillations can be observed below T$_{\text{CDW}}$ as seen from the complex oscillation pattern (Appendix Fig. \ref{appendix_fig1}a). The sign of the initial reflectivity change $\Delta$R/R(t = 0) also flips across T$_{\text{CDW}}$, similar to CsV$_3$Sb$_5$\cite{ratcliff2021coherent}, which is likely due to band renormalization and partial bandgap opening at the CDW transition\cite{liu2021charge, wenzel2022optical}. To better understand the evolution of coherent phonon modes as a function of temperature, we perform a Fourier transformation on the oscillation parts of $\Delta$R/R traces after subtracting a decaying background  to reveal the phonon spectra and plot it in Fig. \ref{fig2}b. There is only one mode centered at 4.0 THz above T$_{\text{CDW}}$. This mode agrees well with the DFT-calculated $A_{1g}$ mode frequency in the pristine phase of RbV$_3$Sb$_5$ (Table \ref{tab:table-1}) and persists in the CDW phase. In contrast, multiple phonon modes emerge in the CDW phase. At T = 5 K, there exists another two conspicuous modes centered at 1.53 and 3.12 THz, along with a 3.75 THz mode and a much weaker mode at 1.75 
THz. Inset Fig. \ref{fig2}c highlights the peak at 1.75 THz. No more modes are detected up to 15 THz. 
The 1.53 THz mode persists during warming but abruptly vanishes at T $\approx$ 100 K, matching with the reported T$_{\text{CDW}}$ = 102 K in RbV$_3$Sb$_5$\cite{yin2021superconductivity, li2021observation}. We note modest local laser heating may cause a slight decrease of the measured T$_{\text{CDW}}$ in the current optical method. No obvious frequency softening is observed for the 1.53 THz mode 
when elevating the temperature, more consistent with a zone-folded phonon mode arising from the CDW order which 
explains the disappearance of this mode in the coherent phonon spectrum above T$_{\text{CDW}}. $ 
Contrary to the single peak feature of the 1.53 THz mode, the 3.12 THz mode is accompanied by a weaker peak at 2.92 THz. This close phonon pair weakens in amplitude as temperature increases and disappears at $\approx$ 85 K. This is reminiscent of a mode in CsV$_3$Sb$_5$ at a similar frequency of 3.1 THz which vanishes in pump-probe phonon spectrum at $\approx$ 60 K, 30 K below T$_{\text{CDW}}$\cite{wang2021unconventional, ratcliff2021coherent}. 

The observations are qualitatively similar in KV$_3$Sb$_5$ (Fig. \ref{fig3}). The sign of the $\Delta$R/R curves also becomes opposite across T$_{\text{CDW}}$, which may likewise arise from the CDW-induced band renormalization and partial CDW gap opening observed by ARPES\cite{luo2022electronic, uykur2022optical}. A zoomed-in $\Delta$R/R curve also highlights multiple phonons in the CDW phase (Appendix Fig. \ref{appendix_fig1}b). Regarding the coherent phonon spectrum, we observe three main peaks at 1.94, 3.04 and 4.00 THz along with weaker peaks at 1.68, 3.45 and 3.64 THz. Inset Fig. \ref{fig3}c highlights the peak at 1.68 THz. The 1.94 THz mode shows minimal frequency softening and is present up to $\sim$75 K in accordance with T$_{\text{CDW}}$ = 78 K\cite{ortiz2021superconductivity}, agreeing with a CDW zone-folded mode. Analogous to RbV$_3$Sb$_5$, the 3.04 THz mode is also accompanied by a weaker peak at 2.71 THz which vanishes at $\approx$ 38 K. 
Above T$_{\text{CDW}}$, only the 4.00 THz mode exists, which matches with the $A_{1g}$ mode in the pristine KV$_3$Sb$_5$ phase (Table \ref{tab:table-2}).

To explain the observed phonon spectrum and determine the CDW structure of both compounds, we perform DFT calculations of phonon frequencies and compare with the phonon spectra detected by TR-reflectivity. In time-resolved reflectivity measurements, the femtosecond pump pulse selectively excites Raman-active phonons coherently\cite{merlin1997generating, stevens2002coherent}. 
The generated coherent phonons by pump pulse then modulates the refractive index by the ion motion\cite{shen1984principles, shen1965theory, giordmaine1966light} which causes the observed transient reflectivity changes. 

\begin{figure*}
    \centering
   \includegraphics[width=17.2cm]{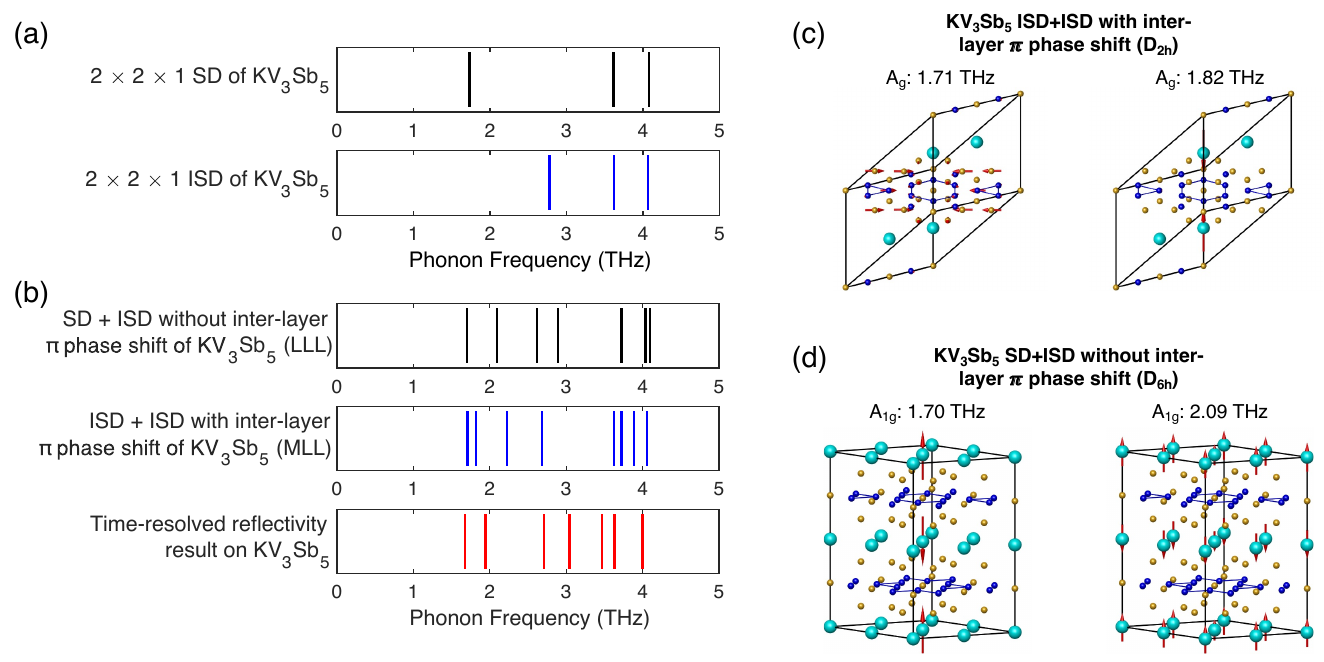}
    \caption{\textbf{Comparison of the phonon spectrum of KV$_3$Sb$_5$ calculated by DFT and measured by time-resolved reflectivity at T = 5 K. } (a) DFT-calculated $A_{1g}$ phonon frequencies in the SD and ISD CDW phases in KV$_3$Sb$_5$. (b) The calculated $A_{1g}$ ($A_g$) Raman mode frequencies in the SD + ISD without interlayer $\pi$ phase shift (ISD+ISD with interlayer $\pi$ phase shift) state and comparison with the time-resolved reflectivity results. The vertical lines in both figures denote the frequency of the measured phonon peak or the fully-symmetric $A_{1g}$ ($A_g$) modes in DFT calculation. (c) The DFT-calculated oscillation pattern of the fully-symmetric Raman active modes in ISD + ISD with interlayer $\pi$ phase shift (MLL) CDW state in KV$_3$Sb$_5$ that are near our detected 1.68 and 1.94 THz modes via time-resolved reflectivity, respectively. The 1.71 and 1.82 THz $A_g$ modes are the lowest and second-lowest frequency mode in the $A_g$ phonon spectrum of MLL phase in (b). (d) The DFT-calculated oscillation pattern of the fully-symmetric Raman active modes in SD + ISD without interlayer $\pi$ phase shift (LLL) CDW state in KV$_3$Sb$_5$ that are near our detected 1.68 and 1.94 THz modes via time-resolved reflectivity, respectively. The 1.70 and 2.09 THz $A_{1g}$ modes are the lowest and second-lowest frequency mode in the $A_{1g}$ phonon spectrum of LLL phase in (b). In (c) and (d), the K atoms are shown in cyan, the V atoms are shown in blue, and the Sb atoms are shown in yellow.} 
    \label{fig5}
\end{figure*}

Our observation is consistent with the displacive excitation of coherent phonons (DECP) mechanism where absorption at the pump frequency is required and only fully symmetric phonon modes are observed\cite{zeiger1992theory, cheng1991mechanism}. Fully symmetric phonon modes host $\Gamma_1^{+}$ symmetry, such as $A_{1g}$ modes in $D_{6h}$ point group and $A_g$ modes in $D_{2h}$ point group. In DECP, the instantaneous electronic excitation caused by absorption of the pump pulse leads to a sudden change in the free energy of the system, which causes the displacement of quasi-equilibrium nuclear coordinates within a unit cell. Such an ultrafast change of quasi-equilibrium position generates a restoring force for the coherent atomic motion of the phonons\cite{zeiger1992theory, lobad2001coherent}. Our pump light wavelength is much larger than the partial CDW gap opening in both compounds\cite{liu2021charge, luo2022electronic, zhou2023electronic, uykur2022optical, wenzel2022optical}, indicating strong absorption. Meanwhile, above T$_{\text{CDW}}$, Raman spectroscopy detected one $A_{1g}$ mode at 4.1 THz and one $E_{2g}$ mode at 3.6 THz with comparable amplitude\cite{wu2022charge} in both (Rb, K)V$_3$Sb$_5$. This 3.6 THz mode is active in pristine phase. This Raman measurement also showed this dominant $E_{2g}$ mode at 3.6 THz persists below T$_{\text{CDW}}$ and has the maximum amplitude among all non-fully-symmetric modes. However, in our time-resolved reflectivity measurements, we only observe the $A_{1g}$ mode near 4.0 THz at temperatures above T$_{\text{CDW}}$ (Table \ref{tab:table-1}, \ref{tab:table-2}). Below T$_{\text{CDW}}$, we do not observe this dominant 3.6 THz $E_{2g}$ mode either. Since we do not observe a mode that shows similar frequency to the 3.6 THz $E_{2g}$ mode and is present at all temperatures, we can rule out the detection of non-fully-symmetric modes in our time-resolved reflectivity measurements. These observations serve as the evidence of DECP, and we assume all the detected phonon modes in our time-resolved reflectivity measurements are fully symmetric modes. This also matches with a previous pump-probe study on CsV$_3$Sb$_5$\cite{ratcliff2021coherent} where DECP was also used to interpret the coherent phonon spectroscopy data.

\begin{table*}[]
\centering
\small
\resizebox{14cm}{!}{%
\begin{tabular}{l|llllllll}
\hline\hline
Pristine RbV$_3$Sb$_5$                                                                                  & \multicolumn{1}{l|}{\begin{tabular}[c]{@{}l@{}}$A_{1g}$ \\ 4.07\end{tabular}} & \begin{tabular}[c]{@{}l@{}}$E_{2g}$ \\ 3.86\end{tabular} &                                                      &                                                      &                                                      &                                                      &                                                      &                                                      \\ \hline
\begin{tabular}[c]{@{}l@{}}2 $\times$ 2 $\times$ 1 SD in RbV$_3$Sb$_5$\end{tabular}                      & \begin{tabular}[c]{@{}l@{}}$A_{1g}$\\ 2.53\end{tabular} & \begin{tabular}[c]{@{}l@{}}$A_{1g}$\\ 3.70\end{tabular} & \multicolumn{1}{l|}{\begin{tabular}[c]{@{}l@{}}$A_{1g}$\\ 4.06\end{tabular}} &  \begin{tabular}[c]{@{}l@{}}$E_{2g}$\\ 0.88\end{tabular}   & \begin{tabular}[c]{@{}l@{}}$E_{2g}$\\ 1.59\end{tabular} &            \begin{tabular}[c]{@{}l@{}}$E_{2g}$\\ 3.54\end{tabular}   &  \begin{tabular}[c]{@{}l@{}}$E_{2g}$\\ 3.86\end{tabular}  &                                                      \\ \hline
\begin{tabular}[c]{@{}l@{}}2 $\times$ 2 $\times$ 1 ISD in RbV$_3$Sb$_5$\end{tabular}                      & \begin{tabular}[c]{@{}l@{}}$A_{1g}$~~~~\\ 2.87\end{tabular} & \begin{tabular}[c]{@{}l@{}}$A_{1g}$~~~~\\ 3.70\end{tabular} & \multicolumn{1}{l|}{\begin{tabular}[c]{@{}l@{}}$A_{1g}$~~~~\\ 4.11\end{tabular}} & \begin{tabular}[c]{@{}l@{}}$E_{2g}$~~~~\\ 1.66\end{tabular}    &  \begin{tabular}[c]{@{}l@{}}$E_{2g}$~~~~\\ 2.62\end{tabular}  &            \begin{tabular}[c]{@{}l@{}}$E_{2g}$~~~~\\ 3.69\end{tabular} & \begin{tabular}[c]{@{}l@{}}$E_{2g}$~~~~\\ 3.87\end{tabular} &                                                      \\ \hline
\begin{tabular}[c]{@{}l@{}}ISD + ISD with inter-layer \\ $\pi$ phase shift of RbV$_3$Sb$_5$\end{tabular}  & \begin{tabular}[c]{@{}l@{}}$A_g$\\ 1.42\end{tabular}  & \begin{tabular}[c]{@{}l@{}}$A_g$\\ 1.60\end{tabular}  & \begin{tabular}[c]{@{}l@{}}$A_g$\\ 2.69\end{tabular}  & \begin{tabular}[c]{@{}l@{}}$A_g$\\ 3.11\end{tabular}  & \begin{tabular}[c]{@{}l@{}}$A_g$\\ 3.65\end{tabular}  & \begin{tabular}[c]{@{}l@{}}$A_g$ \\ 3.76\end{tabular} & \begin{tabular}[c]{@{}l@{}}$A_g$\\ 3.87\end{tabular}  & \begin{tabular}[c]{@{}l@{}}$A_g$~~~~\\ 4.02\end{tabular}  \\ \hline
\begin{tabular}[c]{@{}l@{}}SD + ISD without inter-layer~~~\\ $\pi$ phase shift of RbV$_3$Sb$_5$\end{tabular} & \begin{tabular}[c]{@{}l@{}}$A_{1g}$\\ 1.27\end{tabular} & \begin{tabular}[c]{@{}l@{}}$A_{1g}$\\ 1.78\end{tabular} & \begin{tabular}[c]{@{}l@{}}$A_{1g}$\\ 2.77\end{tabular} & \begin{tabular}[c]{@{}l@{}}$A_{1g}$\\ 3.12\end{tabular} & \begin{tabular}[c]{@{}l@{}}$A_{1g}$\\ 3.72\end{tabular} & \begin{tabular}[c]{@{}l@{}}$A_{1g}$\\ 3.83\end{tabular} & \begin{tabular}[c]{@{}l@{}}$A_{1g}$\\ 3.98\end{tabular} & \begin{tabular}[c]{@{}l@{}}$A_{1g}$\\ 4.10\end{tabular} \\ \hline\hline
\end{tabular}%
}
\caption{Frequency (unit: THz) of the selected Raman-active modes in the pristine and CDW phases of RbV$_3$Sb$_5$ calculated by DFT. Only relevant $A_{1g}$, $E_{2g}$ and $A_g$ modes with frequencies below 4.5 THz are included. }
\label{tab:table-1}
\end{table*}

\begin{table*}[]
\centering
\small
\resizebox{14cm}{!}{%
\begin{tabular}{l|llllllll}
\hline\hline
Pristine KV$_3$Sb$_5$                                                                                  & \multicolumn{1}{l|}{\begin{tabular}[c]{@{}l@{}}$A_{1g}$ \\ 4.05\end{tabular}} & \begin{tabular}[c]{@{}l@{}}$E_{2g}$ \\ 3.85\end{tabular} &                                                      &                                                      &                                                      &                                                      &                                                      &                                                      \\ \hline
2 $\times$ 2 $\times$ 1 SD in KV$_3$Sb$_5$                                                                 & \begin{tabular}[c]{@{}l@{}}$A_{1g}$~~~~\\ 1.73\end{tabular}  & \begin{tabular}[c]{@{}l@{}}$A_{1g}$~~~~\\ 3.61\end{tabular}  & \multicolumn{1}{l|}{\begin{tabular}[c]{@{}l@{}}$A_{1g}$~~~~\\ 4.08\end{tabular}} & \begin{tabular}[c]{@{}l@{}}$E_{2g}$~~~~\\ 0.75\end{tabular} & \begin{tabular}[c]{@{}l@{}}$E_{2g}$~~~~\\ 1.56\end{tabular}  &  \begin{tabular}[c]{@{}l@{}}$E_{2g}$~~~~\\ 3.53\end{tabular}  & \begin{tabular}[c]{@{}l@{}}$E_{2g}$~~~~\\ 3.87\end{tabular}  &                                                      \\ \hline
2 $\times$ 2 $\times$ 1 ISD in KV$_3$Sb$_5$                                                                & \begin{tabular}[c]{@{}l@{}}$A_{1g}$\\ 2.78\end{tabular}  & \begin{tabular}[c]{@{}l@{}}$A_{1g}$\\ 3.62\end{tabular}  & \multicolumn{1}{l|}{\begin{tabular}[c]{@{}l@{}}$A_{1g}$\\ 4.06\end{tabular}} & \begin{tabular}[c]{@{}l@{}}$E_{2g}$\\ 1.61\end{tabular} & \begin{tabular}[c]{@{}l@{}}$E_{2g}$\\ 2.57\end{tabular} & \begin{tabular}[c]{@{}l@{}}$E_{2g}$\\ 3.64\end{tabular} & \begin{tabular}[c]{@{}l@{}}$E_{2g}$\\ 3.86\end{tabular}   &                                                      \\ \hline
\begin{tabular}[c]{@{}l@{}}ISD + ISD with inter-layer \\ $\pi$ phase shift of KV$_3$Sb$_5$\end{tabular}  & \begin{tabular}[c]{@{}l@{}}$A_g$\\ 1.71\end{tabular}   & \begin{tabular}[c]{@{}l@{}}$A_g$\\ 1.82\end{tabular}   & \begin{tabular}[c]{@{}l@{}}$A_g$\\ 2.22\end{tabular}  & \begin{tabular}[c]{@{}l@{}}$A_g$\\ 2.69\end{tabular}  & \begin{tabular}[c]{@{}l@{}}$A_g$\\ 3.63\end{tabular}  & \begin{tabular}[c]{@{}l@{}}$A_g$\\ 3.72\end{tabular}  & \begin{tabular}[c]{@{}l@{}}$A_g$\\ 3.88\end{tabular}  & \begin{tabular}[c]{@{}l@{}}$A_g$~~~~\\ 4.06\end{tabular}  \\ \hline
\begin{tabular}[c]{@{}l@{}}SD + ISD without inter-layer~~~\\ $\pi$ phase shift of KV$_3$Sb$_5$\end{tabular} & \begin{tabular}[c]{@{}l@{}}$A_{1g}$\\ 1.70\end{tabular}  & \begin{tabular}[c]{@{}l@{}}$A_{1g}$\\ 2.09\end{tabular}  & \begin{tabular}[c]{@{}l@{}}$A_{1g}$\\ 2.62\end{tabular} & \begin{tabular}[c]{@{}l@{}}$A_{1g}$\\ 2.89\end{tabular} & \begin{tabular}[c]{@{}l@{}}$A_{1g}$\\ 3.72\end{tabular} & \begin{tabular}[c]{@{}l@{}}$A_{1g}$\\ 3.73\end{tabular} & \begin{tabular}[c]{@{}l@{}}$A_{1g}$\\ 4.04\end{tabular} & \begin{tabular}[c]{@{}l@{}}$A_{1g}$\\ 4.10\end{tabular} \\ \hline\hline
\end{tabular}%
}
\caption{Frequency (unit: THz) of the selected Raman-active modes in the pristine and CDW phases of KV$_3$Sb$_5$ calculated by DFT. Only relevant $A_{1g}$, $E_{2g}$ and $A_g$ modes with frequencies below 4.5 THz are included. }
\label{tab:table-2}
\end{table*}

\begin{figure}[t]
    \centering
   \includegraphics[width=8.6cm]{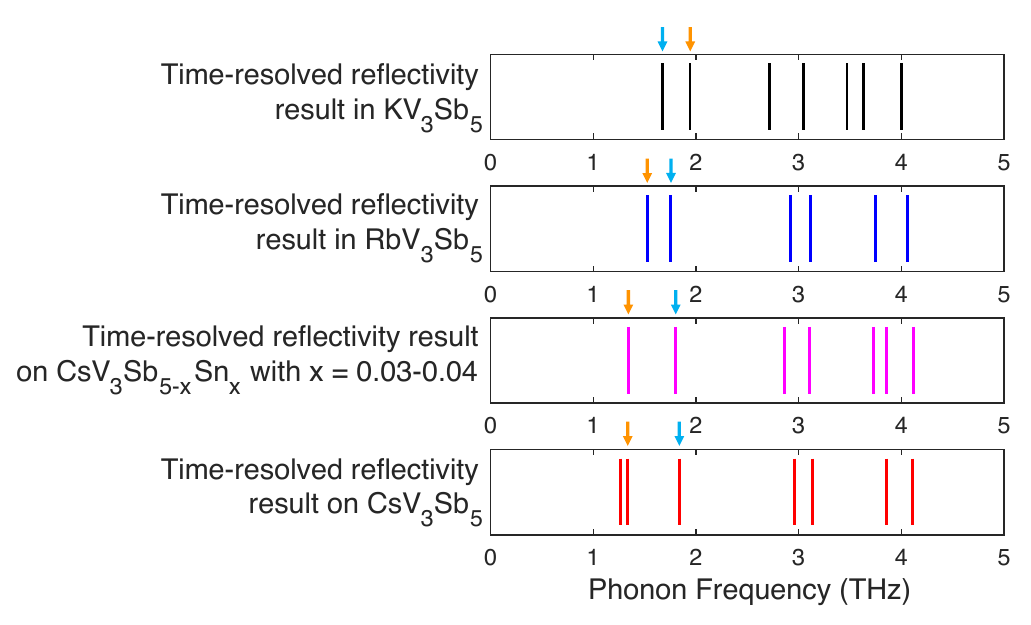}
    \caption{The measured coherent phonon spectrum by time-resolved reflectivity in KV$_3$Sb$_5$, RbV$_3$Sb$_5$, CsV$_3$Sb$_5$, and Sn-doped CsV$_3$Sb$_5$. The spectrum for Sn-doped CsV$_3$Sb$_5$ is taken at T = 3.5 K, while the spectra for other 3 compounds are taken at T = 5 K. The data from CsV$_3$Sb$_5$ and Sn-doped CsV$_3$Sb$_5$ will be published elsewhere\cite{deng2024}. The orange arrow marks the evolution of the frequency of the alkali metal atom mode. The blue arrow marks the evolution of the frequency of the V and Sb atom related mode near 1.7 - 1.8 THz. } 
    \label{fig6}
\end{figure}

We first analyze the possible CDW structure in RbV$_3$Sb$_5$. Fig. \ref{fig4}a shows the calculated $A_{1g}$ modes in the two $C_6$-symmetric 2 $\times$ 2 $\times$ 1 CDW phases, Star-of-David (SD) and inverse star-of-David (ISD), in the relevant frequency range. Both phases cannot explain the observed phonon spectra due to the lack of the two lower frequency modes and two close modes near 3.12 THz. Thus, interlayer modulation of the CDW order along c axis must be included. Regarding the 2 $\times$ 2 $\times$ 2 CDW phases, we consider the two reported structures: ISD + ISD with interlayer $\pi$ phase shift (MLL) with broken $C_6$, and SD + ISD without interlayer $\pi$ phase shift (LLL) with preserved $C_6$. The calculated fully-symmetric phonon frequencies for both structures have been listed in Table \ref{tab:table-1} and Fig. \ref{fig4}b. According to Fig. \ref{fig4}b, both host two fully symmetric modes between 1 and 2 THz, while the frequency difference of these two modes in MLL state are better matched with the detected 1.53 and 1.75 THz mode. Further supportive evidence of MLL arises from the DFT-calculated phonon oscillation patterns (Fig. \ref{fig4}c, d). As shown by DFT, for the two $A_g$ modes near the observed phonon frequencies in MLL, the 1.42 THz mode is Rb-related while the 1.60 THz mode involves V and Sb atoms. If MLL is the actual CDW structure in RbV$_3$Sb$_5$, our observed 1.53 THz mode should correspond to the calculated 1.42 THz $A_g$ mode that is related with Rb atoms, and our observed 1.75 THz mode should correspond to the calculated 1.60 THz $A_g$ mode that is related with V and Sb atoms (Fig. \ref{fig4}c). When moving from Rb to Cs, the frequency of the 1.53 THz mode will decrease and the frequency of the 1.75 THz mode will remain similar. This matches with the phonon spectrum of CsV$_3$Sb$_{5-\text{x}}$Sn$_\text{x}$ with x = 0.03-0.04 which also has an MLL-type CDW\cite{kang2022charge, deng2024}(Fig. \ref{fig6}). Our measurement on CsV$_3$Sb$_{5-\text{x}}$Sn$_\text{x}$ with x = 0.03-0.04 shows a 1.34 THz mode that corresponds to a Cs related mode, and a 1.80 THz mode that corresponds to a V and Sb related mode\cite{deng2024}. When switching Rb to Cs, the 1.53 THz Rb-related mode in RbV$_3$Sb$_5$ decreases in frequency and evolves to the 1.34 THz Cs-related mode in this Sn-doped CsV$_3$Sb$_5$. Meanwhile, the frequency of the 1.75 THz V and Sb-related mode in RbV$_3$Sb$_5$ should remain similar and evolves to the 1.80 THz V and Sb-related mode in this Sn-doped CsV$_3$Sb$_5$. 
We can also rule out the LLL phase since both phonons below 2 THz in LLL are Rb-related only (Appendix A). We thus conclude the ISD + ISD with interlayer $\pi$ phase shift (MLL) is the actual CDW structure in RbV$_3$Sb$_5$, in which six-fold rotational symmetry is broken. Also, in our previous study\cite{xu2022three}, we observed three birefringence domains on both compounds and their principal axes are 120$^{\circ}$ relative to each other, showing strong evidence of $C_6$ breaking as well. Since the reported birefringence domain size is $\approx$100 $\mu$m and the laser spot size in our time-resolved reflectivity experiments is $\approx$10 $\mu$m, we can measure in a single birefringence domain. This allows us to pinpoint $C_6$-breaking MLL as the exact CDW structure. We note that while there are more fully-symmetric modes in calculation than observed, some modes could be missing in our observation due to their weak modulations on refractive index and reflectivity.

We then perform similar phonon frequency calculations on KV$_3$Sb$_5$. Fig. \ref{fig5}a compares the calculated $A_{1g}$ phonons of SD and ISD with the time-resolved reflectivity results. Clearly, these two 2 $\times$ 2 $\times$ 1 CDW structures fail to explain the two close modes near 3.04 THz and the multiple observed phonons near 3.5 THz. We again compare our measured phonon spectrum with MLL and LLL. The calculated fully-symmetric phonon frequencies for both structures have been listed in Table \ref{tab:table-2} and Fig. \ref{fig5}b. For the phonon spectrum of LLL phase of KV$_3$Sb$_5$, it shows four phonons at 3.72, 3.73, 4.04 and 4.10 THz. The frequency differences of some adjacent phonons (0.01 THz, 0.06 THz) are too small to match well with our measured values. Comparatively, the frequency values and frequency difference of the observed multiple modes near 3.5 THz (i.e. the observed 3.45 and 3.64 THz modes) match better with the MLL phase. For the modes below 2 THz, we note that a previous Raman study\cite{wu2022charge} observed a 51 cm$^{-1}$ = 1.53 THz fully-symmetric mode in RbV$_3$Sb$_5$ and a 64 cm$^{-1}$ = 1.92 THz fully symmetric mode in KV$_3$Sb$_5$, matching with our observed 1.53 THz mode in RbV$_3$Sb$_5$ and 1.94 THz mode in KV$_3$Sb$_5$ respectively. This Raman study also established the frequency evolution from 1.53 THz to 1.94 THz when changing Rb to K. If MLL is the actual CDW structure in KV$_3$Sb$_5$, our observed 1.68 THz mode should correspond to the calculated 1.71 THz $A_g$ mode that is related with V and Sb atoms, and our observed 1.94 THz mode should correspond to the calculated 1.82 THz $A_g$ mode that is related with K atoms (Fig. \ref{fig5}c). Considering the CDW structure of RbV$_3$Sb$_5$ is determined to be MLL, by switching from Rb to K, one expects the frequency of the Rb-related 1.53 THz mode in RbV$_3$Sb$_5$ to increase and the frequency of the V- and Sb-related 1.75 THz mode in RbV$_3$Sb$_5$ to be similar. This exactly agrees with our observed 1.94 and 1.68 THz mode in KV$_3$Sb$_5$ respectively: the 1.53 THz mode in RbV$_3$Sb$_5$ evolves to 1.94 THz mode in KV$_3$Sb$_5$, and the 1.75 THz mode in RbV$_3$Sb$_5$ evolves to 1.68 THz mode in KV$_3$Sb$_5$, indicating the CDW structure in KV$_3$Sb$_5$ is MLL as well. Analogous argument as for RbV$_3$Sb$_5$ can rule out LLL (Appendix A). We thus conclude the CDW structure in KV$_3$Sb$_5$ is also ISD + ISD with interlayer $\pi$ phase shift.

\begin{table}[]
\centering
\resizebox{8.7cm}{!}{%
\begin{tabular}{c|c|c}
\hline
\begin{tabular}[c]{@{}c@{}}2 $\times$ 2 $\times$ 1 ISD of \\ RbV$_3$Sb$_5$ ($D_{6h}$)\end{tabular} &                                & \begin{tabular}[c]{@{}c@{}}2 $\times$ 2 $\times$ 2 ISD+ISD with inter-layer \\ $\pi$ phase shift of RbV$_3$Sb$_5$ ($D_{2h}$)\end{tabular} \\ \hline
\multirow{2}{*}{$E_{2g}$:  1.66 THz}                                                             & \multirow{2}{*}{$\rightarrow$} & $A_g$:  1.60 THz                                                                                                                        \\  
                                                                                                   &                                & $B_{1g}$:  1.66 THz                                                                                                                     \\ \hline
\end{tabular}%
}
\end{table}

\begin{table}[t]
\centering
\resizebox{8.6cm}{!}{%
\begin{tabular}{c|c|c}
\hline
\begin{tabular}[c]{@{}c@{}}2 $\times$ 2 $\times$ 1 ISD of \\ RbV$_3$Sb$_5$ ($D_{6h}$)\end{tabular} &                                & \begin{tabular}[c]{@{}c@{}}2 $\times$ 2 $\times$ 2 ISD+ISD with inter-layer \\ $\pi$ phase shift of RbV$_3$Sb$_5$ ($D_{2h}$)\end{tabular} \\ \hline
$A_{1g}$:  2.87 THz                                                                    & $\rightarrow$                  & $A_g$:  3.11 THz                                                                                                            \\ \hline
\multirow{2}{*}{$E_{2g}$:  2.62 THz}                                                   & \multirow{2}{*}{$\rightarrow$} & $A_g$:  2.69 THz                                                                                                            \\ 
                                                                                          &                                & $B_{1g}$:  2.82 THz                                                                                                           \\ \hline
\end{tabular}%
}
\caption{DFT-calculated phonon split due to six-fold rotational symmetry breaking in RbV$_3$Sb$_5$. The pair of $A_g$ modes near 3.1 THz shown in the second table correspond to the observed two close modes near 3.1 THz. }
\label{tab:table-3}
\end{table}

\begin{table}[t]
\centering
\resizebox{8.6cm}{!}{%
\begin{tabular}{c|c|c}
\hline
\begin{tabular}[c]{@{}c@{}}2 $\times$ 2 $\times$ 1 ISD of\\ KV$_3$Sb$_5$ ($D_{6h}$)\end{tabular} &                                & \begin{tabular}[c]{@{}c@{}}2 $\times$ 2 $\times$ 2 ISD+ISD with inter-layer \\ $\pi$ phase shift of KV$_3$Sb$_5$ ($D_{2h}$)\end{tabular} \\ \hline
\multirow{2}{*}{$E_{2g}$:  1.61 THz}                                                        & \multirow{2}{*}{$\rightarrow$} & $A_g$:  1.71 THz                                                                                                                                             \\ 
                                                                                              &                                & $B_{1g}$:  1.65 THz                                                                                                                                          \\ \hline
\end{tabular}%
}
\end{table}

\begin{table}[htbp]
\centering
\resizebox{8.7cm}{!}{%
\begin{tabular}{c|c|c}
\hline
\begin{tabular}[c]{@{}c@{}}2 $\times$ 2 $\times$ 1 ISD of \\KV$_3$Sb$_5$ ($D_{6h}$)\end{tabular} &                                & \begin{tabular}[c]{@{}c@{}}2 $\times$ 2 $\times$ 2 ISD+ISD with inter-layer \\ $\pi$ phase shift of KV$_3$Sb$_5$ ($D_{2h}$)\end{tabular} \\ \hline
$A_{1g}$:  2.78 THz                                                             & $\rightarrow$                  & $A_g$:  2.69 THz                                                                                                                                     \\ \hline
\multirow{2}{*}{$E_{2g}$:  2.57 THz}                                            & \multirow{2}{*}{$\rightarrow$} & $A_g$:  2.22 THz                                                                                                                                     \\ 
                                                                                   &                                & $B_{1g}$:  2.45 THz                                                                                                                                    \\ \hline
\end{tabular}%
}
\caption{DFT-calculated phonon split due to six-fold rotational symmetry breaking in KV$_3$Sb$_5$. The pair of $A_g$ modes near 3 THz shown in the second table correspond to the observed two close modes near 3.0 THz. }
\label{tab:table-4}
\end{table}

We note a previous Raman spectroscopy study on AV$_3$Sb$_5$ family also revealed the two close modes near 3 THz for both RbV$_3$Sb$_5$ and KV$_3$Sb$_5$ and matched well with our measured frequencies\cite{wu2022charge}. Therein, a 2 $\times$ 2 $\times$ 4 CDW order has been shown to give rise to $A_{1g}$ doublets and was utilized to explain these detected $A_{1g}$ pairs. However, XRD measurements on either RbV$_3$Sb$_5$ or KV$_3$Sb$_5$ did not find any evidence of 2 $\times$ 2 $\times$ 4 orders\cite{kautzsch2023structural, scagnoli2024resonant}. Thus, we suggest 2 $\times$ 2 $\times$ 4 CDW orders may not be the reason for our observed two close modes near 3 THz in (Rb, K)V$_3$Sb$_5$. Instead, the $C_6$-breaking MLL order can explain the observed two close modes near 3 THz. 


Our measurements of the coherent phonon dynamics in Rb- and KV$_3$Sb$_5$ via time-resolved reflectivity provide a new perspective to resolve the dispute of the CDW structure in these compounds. Although in-plane 2 $\times$ 2 $\times$ 1 CDW orders were proposed by some ARPES\cite{kato2022three} and NMR\cite{zhang2024nmr} measurements, the observation of the two close modes near 3 THz and the alkali-metal modes below 2 THz in Rb- and KV$_3$Sb$_5$ unambiguously rules out the two 2 $\times$ 2 $\times$ 1 CDW structures for both compounds. The observation of the 1.53 and 1.75 THz modes in RbV$_3$Sb$_5$ along with 1.94 and 1.68 THz modes 
in KV$_3$Sb$_5$ also supports the $C_6$-breaking MLL phase against LLL phase in both materials, in agreement with previous theoretical predictions that MLL is energetically more favorable than LLL\cite{tan2021charge, ratcliff2021coherent} and sharing the result of recent ARPES\cite{jiang2023observation, kang2022charge} and NMR\cite{wang2023structure, frassineti2023microscopic} measurements. In addition to previous STM studies\cite{li2022rotation, xing2024optical} visualizing $C_6$ breaking at low temperatures from different CDW peak height along three in-plane directions, the persistence of the observed 1.53 and 1.94 THz modes up to T$_{\text{CDW}}$ in our time-resolved reflectivity results confirms the existence of MLL phase and six-fold rotational symmetry breaking up to T$_{\text{CDW}}$ in both Rb- and KV$_3$Sb$_5$. 

The collation of the measured phonon spectrum in all AV$_3$Sb$_5$ members from Cs to K by time-resolved reflectivity allows us to track the evolution of the phonon frequencies (Fig. \ref{fig6}). 
We observe prominent long-lived phonon modes at 1.33 THz in CsV$_3$Sb$_5$\cite{deng2024}, 1.53 THz in RbV$_3$Sb$_5$ and 1.94 THz in KV$_3$Sb$_5$, all of which vanish at T$_{\text{CDW}}$. These modes match with alkali metal atom related phonon modes seen by DFT. As the atomic mass decreases from Cs, Rb to K, we expect the phonon vibration frequency to increase. This explains the detected phonon frequency evolution from 1.33 THz in CsV$_3$Sb$_5$ to 1.53 THz in RbV$_3$Sb$_5$ to 1.94 THz in KV$_3$Sb$_5$ (orange arrow in Fig. \ref{fig6}). Our time-resolved reflectivity measurements also observe a weaker fully-symmetric $A_g$ mode near 1.7 - 1.8 THz in all AV$_3$Sb$_5$ members in Fig. \ref{fig6} that is not observed in previous pump-probe measurements. This mode is related to V and Sb atoms, and whose frequency only shows slight decreases when the alkali metal atom mass decreases from Cs, Rb to K, i.e. from 1.8 THz in CsV$_3$Sb$_5$ to 1.75 THz in RbV$_3$Sb$_5$ to 1.68 THz in KV$_3$Sb$_5$ (blue arrow in Fig. \ref{fig6}). This mode comes from $C_6$ symmetry breaking as it can be explained by splitting of an $E_{2g}$ mode ($E_{2g}$ $\rightarrow$ $A_g + B_{1g}$) when $C_6$ is reduced to $C_2$ (Table \ref{tab:table-3} \& \ref{tab:table-4})\cite{deng2024}. These parent $E_{2g}$ modes have been detected by Raman\cite{wu2022charge} and shown to evolve from 1.83 THz in CsV$_3$Sb$_5$ to 1.77 THz in RbV$_3$Sb$_5$ to 1.68 THz in KV$_3$Sb$_5$, matching with our measured frequency values and evolution. In addition, in all compounds, we observe a close pair of fully-symmetric phonons near 3 THz. Previous pump-probe studies on CsV$_3$Sb$_5$\cite{wang2021unconventional, ratcliff2021coherent} suggested the 3 THz mode to be linked to the 1Q uniaxial order observed by STM below $\approx$ 60 K\cite{zhao2021cascade, chen2021roton}, much lower than T$_{\text{CDW}}$. We note the 1Q uniaxial order is not seen in KV$_3$Sb$_5$\cite{jiang2021unconventional}, thus indicating the 1Q uniaxial order might not be the reason for the observed 3 THz modes. In contrast, we point out the 3Q MLL phase with interlayer $\pi$ phase shift can already explain the two close modes near 3 THz. This close phonon pair near 3 THz is also a direct consequence of $C_6$ symmetry breaking, as the lower frequency branch $A_g$ mode comes from splitting of an $E_{2g}$ mode ($E_{2g}$ $\rightarrow$ $A_g + B_{1g}$) induced by broken $C_6$ (Table \ref{tab:table-3} \& \ref{tab:table-4}).  Moreover, our measurements push the onset temperature of $C_6$ rotational symmetry breaking up to T$_{\text{CDW}}$ in Rb- and KV$_3$Sb$_5$ rather than below T$_{\text{CDW}}$. Previous Raman studies on CsV$_3$Sb$_5$\cite{liu2022observation, he2024anharmonic} showed this 3 THz mode exhibits significant weakening and broadening upon warming toward T$_{\text{CDW}}$, becoming overdamped between 60 and 90 K with the linewidth as large as 50 cm$^{-1}$ right below T$_{\text{CDW}}$. This might explain why the 3 THz modes vanish in our experiment at $\sim$20$-$30 K below T$_{\text{CDW}}$ in RbV$_3$Sb$_5$ and KV$_3$Sb$_5$.

In conclusion, our coherent phonon excitation studies determine the precise structure and symmetry in the CDW phase of RbV$_3$Sb$_5$ and KV$_3$Sb$_5$. We resolve the fully-symmetric phonon spectrum via ultrafast time-resolved reflectivity, from which we pinpoint the microscopic 3D 2 $\times$ 2 $\times$ 2 CDW pattern to be ISD + ISD with interlayer $\pi$ phase shift (MLL) by comparing with DFT calculations, indicating $C_6$ rotational symmetry breaking up to T$_{\text{CDW}}$. Our results thus provide fresh insights into the current understanding of CDW state in AV$_3$Sb$_5$ and essential guidance on the development of microscopic theory of emergent electronic orders stabilized within the CDW phase, such as chiral loop current states and superconductivity. Looking forward, we envision our technique will help deciphering of the evolution of CDW order across the extremely rich phase diagram of AV$_3$Sb$_5$, extending new vision for fundamental studies at the intersection of frustration, topology, and strong correlations.

\section*{ACKNOWLEDGEMENT}
The construction of the pump-probe setup was supported by the Air Force Office of Scientific Research under award no. FA9550-22-1-0410. Q.D. was mainly supported by the Vagelos Institute of Energy Science and Technology graduate fellowship and also partly supported by the Air Force Office of Scientific Research under award no. FA9550-22-1-0410 and the NSF EPM program under grant no. DMR-2213891. S.D.W. and B.R.O. gratefully acknowledge support via the UC Santa Barbara NSF Quantum Foundry funded via the Q-AMASE-i program under award DMR-1906325. B.R.O. thanks support from the U.S. Department of Energy (DOE), Office of Science, Basic Energy Sciences (BES), Materials Sciences and Engineering Division. B.Y. acknowledges the financial support by the Israel Science Foundation (ISF: 2932/21, 2974/23), German Research Foundation (DFG, CRC-183, A02), and by a research grant from the Estate of Gerald Alexander.


\textit{Correspondence: }Correspondence and requests for materials
should be addressed to L.W. (liangwu@sas.upenn.edu)



\clearpage
\newpage

\bibliography{RbV3Sb5_KV3Sb5_TRR_refs.bib}

\clearpage

\setcounter{figure}{0}
\renewcommand{\figurename}{{\bf{Appendix Figure}}}

\section*{APPENDIX A: Ruling out LLL CDW Structure in RbV$_3$Sb$_5$ and KV$_3$Sb$_5$}

The DFT-calculated fully-symmetric phonon modes of different CDW structures in RbV$_3$Sb$_5$ and KV$_3$Sb$_5$ are shown in Table \ref{tab:table-1} and \ref{tab:table-2}. 

\subsection*{I. In RbV$_3$Sb$_5$}

For the DFT-calculated two $A_g$ modes below 2 THz in MLL that are near the observed 1.53 and 1.75 THz modes, the lower frequency mode (1.42 THz) only involves Rb atoms, while the higher frequency mode (1.60 THz) involves V and Sb atoms. For the DFT-calculated two $A_{1g}$ modes below 2 THz in LLL (at 1.27 THz and 1.78 THz) that are near the observed 1.53 and 1.75 THz modes, both only involve Rb atom oscillations. See Fig. \ref{fig4}c, d.

If LLL was the correct CDW structure, both the experimentally detected 1.53 and 1.75 THz modes would correspond to Rb-related modes. When Rb is switched to Cs, their frequencies should decrease. Thus, in CsV$_3$Sb$_5$ with the coexistence of SD and ISD which contains the whole LLL phonon set\cite{deng2024}, one would expect two modes with frequency below 1.53 THz and 1.75 THz respectively, with the higher frequency mode much weaker than the lower frequency mode, as observed in CsV$_3$Sb$_5$\cite{deng2024}. Although the observed 1.26 and 1.33 THz mode in CsV$_3$Sb$_5$ satisfy this frequency evolution\cite{deng2024}, the higher frequency 1.33 THz mode always show stronger amplitude instead of being much weaker than the lower frequency 1.26 THz mode. This rules out the LLL as the CDW structure in RbV$_3$Sb$_5$.

\subsection*{II. In KV$_3$Sb$_5$}

For MLL, DFT calculation shows two $A_g$ modes at 1.71 THz and 1.82 THz that are near the observed 1.68 and 1.94 THz modes. The 1.71 THz mode involves the oscillations of Sb and V atoms, while the 1.82 THz mode only involves the oscillation of K atoms. For LLL, DFT calculation shows two $A_{1g}$ modes at 1.70 THz and 2.09 THz that are near the observed 1.68 and 1.94 THz modes. Both modes only involve the oscillations of K atoms. See Fig. \ref{fig5}c, d.

If LLL was the correct CDW structure, both the experimentally detected 1.94 and 1.68 THz modes would correspond to K-related modes. When K is switched to Cs, their frequencies should decrease. Thus, in CsV$_3$Sb$_5$ with the coexistence of SD and ISD which contains the whole LLL phonon set\cite{deng2024}, one would expect two modes with frequency below 1.68 THz and 1.94 THz respectively, with the higher frequency mode much stronger than the lower frequency mode. Although the observed 1.26 and 1.33 THz mode in CsV$_3$Sb$_5$ satisfy this frequency evolution\cite{deng2024}, the amplitude of these two modes are comparable, in contrast to KV$_3$Sb$_5$ where the amplitude of the higher frequency 1.94 THz mode is more than one order of magnitude stronger than the lower frequency 1.68 THz mode. This rules out the LLL as the CDW structure in KV$_3$Sb$_5$.

\begin{figure*}[]
    \centering
    \includegraphics[width=12cm]{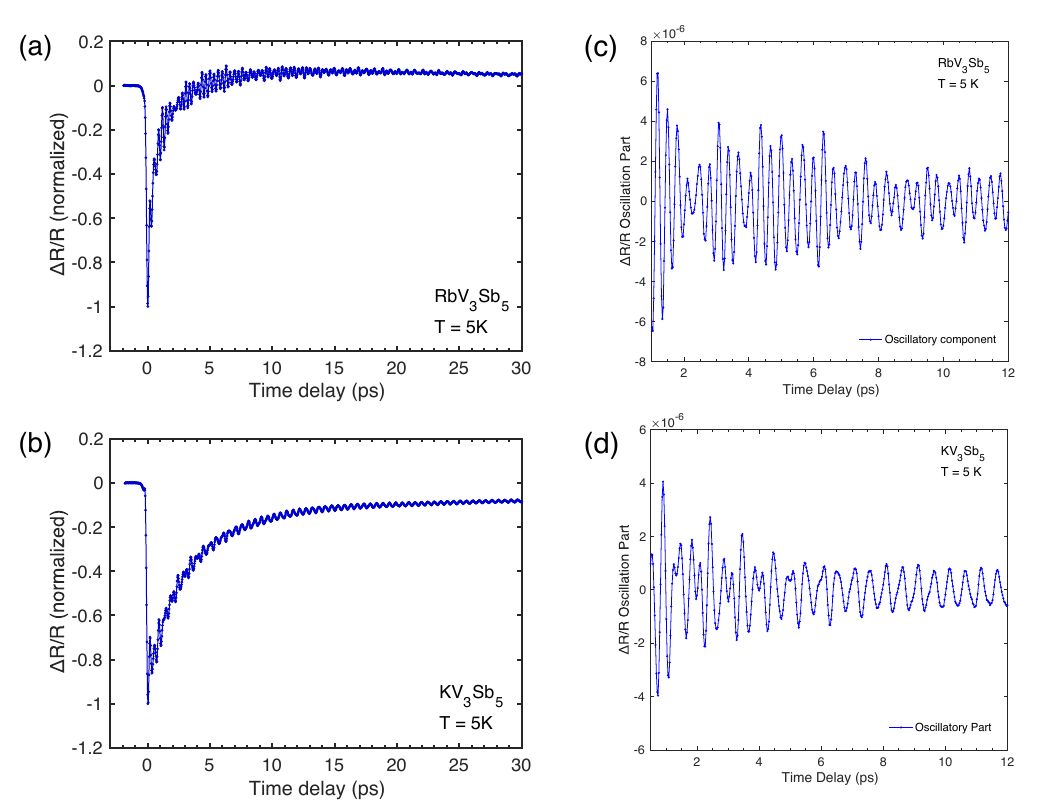}
    \caption{The zoomed-in plots highlighting the oscillation patterns of time-resolved reflectivity curves at 5 K for (a) RbV$_3$Sb$_5$ and (b) KV$_3$Sb$_5$. (c) The oscillatory part of the measured time-resolved reflectivity curve on RbV$_3$Sb$_5$ at 5 K in the first 12 ps. (d) The oscillatory part of the measured time-resolved reflectivity curve on KV$_3$Sb$_5$ at 5 K in the first 12 ps.}
    \label{appendix_fig1}
\end{figure*}

\end{document}